\documentclass[twocolumn]{webofc}
\usepackage[varg]{txfonts}   % Web of Conferences font
\usepackage{hyperref}
\usepackage{url}
\usepackage{graphicx}
\usepackage{subcaption}
\hypersetup{colorlinks=true,citecolor=blue,urlcolor=blue,linkcolor=blue}
\usepackage{physics}
\usepackage{xcolor}

\begin{document}
\title{Exploring Interplays Between $^3\text{P}_2$ Neutron Superfluid Vortices and $^1\text{S}_0$ Proton Fluxtubes in the Outer Core of Neutron Stars}

\author{\firstname{Tatsuhiro} \lastname{Hattori}\inst{1}\fnsep\thanks{\email{hattori.t.a850@m.isct.ac.jp}} \and
        \firstname{Kazuyuki} \lastname{Sekizawa}\inst{1,2,3}\fnsep\thanks{\email{sekizawa@phys.sci.isct.ac.jp}} 
        % etc.
}

\institute{Department of Physics, School of Science, Institute of Science Tokyo, Tokyo 152-8550, Japan
\and
           Nuclear Physics Division, Center for Computational Sciences, University of Tsukuba, Ibaraki 305-8557, Japan
\and
           RIKEN Nishina Center, Saitama 351-0198, Japan
          }

\abstract{
In the outer core of neutron stars, $^3$P$_2$ superfluid neutrons and $^1$S$_0$ superconducting protons are deemed to exist, forming quantum vortices and magnetic fluxtubes, respectively. Those quantum vortices and fluxtubes play an important role in explaining observed sudden changes of rotational frequency, known as pulsar ``glitches.'' While the most of conventional glitch models rely on pinning/unpinning dynamics of neutron $^1\text{S}_0$ superfluid vortices in the inner crust, contributions of the outer core have not been ruled out. However, the latter possibility has been less explored so far and further thorough investigations are desired. In this study, we are thus developing a microscopic model based on spin-2 Gross-Pitaevskii equation (GPE) for neutron $^3$P$_2$ superfluid vortices coupled with Ginzburg-Landau equation (GLE) for fluxtubes associated with superconducting $^1\text{S}_0$ protons. In this contribution, we outline our theoretical framework and report tentative results showing how shape of quantum vortices could be affected by the presence of a proton fluxtube.
}
\maketitle
\section{Introduction}
\label{intro}
% Neutron Star

Neutron stars are massive and dense stellar objects that are formed as a remnant of supernova explosion of massive stars. The interior of neutron stars exhibits various phases depending on density, which are typically classified as four regions: the outer crust, inner crust, outer core, and inner core, from surface to center. The outer crust is made of a Coulomb lattice of nuclei and background relativistic electron gas, where the nuclei become more and more neutron rich as the density increases because of electron capture processes. At a certain critical density ($\sim4\times10^{11}$\,g/cm$^3$), neutrons start dripping out of the nuclei and the Coulomb lattice is immersed in dripped neutrons that form $^1\text{S}_0$ superfluid. As density increases higher, going deeper inside of the inner crust, the distance between nuclei decreases and nuclei eventually merge to form uniform nuclear matter (at $\sim 10^{14}$\,g/cm$^3$), which defines a transition from the inner crust to the outer core. In the outer core region, protons are considered to be type-II $^1\text{S}_0$ superconductor, while neutrons are in $^3\text{P}_2$ superfluid phase \cite{Apj_paul_ray_2019}.

% Glitch
Pulsars are rotating neutron stars, which rotate with a period ranging typically from milliseconds to seconds. From pulsar observations, sudden changes of rotational frequency, known as ``glitches,'' have been discovered. First observation was the glitch of Vela pulsar \cite{nature_vanghan_large_mills,ap_cheng_1988}, since then it has become a typical glitching pulsar that shows glitches quasi-periodically, 21 times in 47 years on Vela pulsar \cite{Xu2016}. Typical glitches are sudden spin-up events that are observed in pulsars that are gradually slowing down due to the electromagnetic radiation. However, recent observations reported other types of glitches: accretion-powered pulsars with a spin-down glitch (\textit{e.g.}, NGC300 ULC-1\cite{ray_apjl_2019}) rotation-powered pulsars with a spin-down glitch (\textit{e.g.}, PSR B0540-69\cite{apjl_youli_tuo_2024}).

% Glitch models
The key to understand the mechanism of the pulsar glitch phenomenon is superfluidity. In rotating superfluid, quantum vortices are formed that carry quantized angular momentum. Thus, superfluid plays a role of angular momentum reservoir which stores angular momentum as quantum vortices. The stored angular momentum could be the resource of glitches and several mechanisms have been proposed to explain observed glitches. In the vortex avalanche model, quantum vortices in the inner crust are pinned to the nuclear Coulomb lattice, and they undergo catastrophic unpinning to trigger glitches \cite{ap_cheng_1988}. This model can explain the spin-up glitches like Vela pulsar's ones, but trigger mechanism of the avalanche is still an open question. In addition, this model cannot describe spin-down glitches in rotation-powered pulsars.

Recently, a vortex network model has been proposed, suggesting that a massive vortex network, formed by connecting neutron $^1\text{S}_0$ superfluid vortices in the inner crust and neutron $^3\text{P}_2$ superfluid vortices in the outer core, as an origin of glitches \cite{sci_rep_marmorinni_nitta_2024}. This model has been successful in describing the size-frequency systematics of observed glitches without invoking pinning and avalanches. On the other hand, spin-down glitches in accretion-powered pulsars and spin-down glitches in rotation-powered pulsars are not well described in the latter model. In addition, the effects of interactions between neutron $^3\text{P}_2$ superfluid vortices and proton fluxtubes are not considered.

% outer core: flux tube-vortex
In the outer core region, $^3\text{P}_2$ superfluid neutrons can form half-integer quantum vortices (HQVs). In addition, $^1\text{S}_0$ superconducting protons quantize magnetic field as fluxtubes. In Ref.~\cite{mnras_thong_melatos_drummond_2023}, the interaction between neutron $^1\text{S}_0$ superfluid vortices and proton fluxtubes are analyzed with a proton-neutron effective interaction. However, as mentioned earlier, neutrons are expected to form $^3\text{P}_2$ superluid in the outer core region and the extension of the work of Ref.~\cite{mnras_thong_melatos_drummond_2023} to $^3\text{P}_2$ vortices is mandatory. $^3\text{P}_2$ vortices can not only be affected by the neutron-proton interaction, but also by a Zeeman like spin megnetic-field interaction. To this end, we are developing a theoretical framework that can describe vortices-fluxtubes configurations microscopically, treating proton-neutron interactions as well as induced magnetic field effects in a self-consistent manner.

% Purpose
In this contribution, we outline our theoretical framework to describe $^3\text{P}_2$ superfluid neutrons in the presence of $^1\text{S}_0$ superconducting protons. We report some selected examples of tentative results on the static behavior of quantum vortices and fluxtubes in the outer core region, taking into account polarization effects induced by spin magnetic-field interactions.

\section{Methods}
\label{sec-1}

% For bibliography use \cite{RefJ}, \cite{RefB}

\subsection{GPE for $^3\text{P}_2$ Neutron Cooper Pairs}
\label{sec-2}

Since the total spin of a $^3\text{P}_2$ Cooper pair is two, neutron $^3\text{P}_2$ superfluidity may be regarded as Bose-Einstein Condensate (BEC) of spin-2 spinor bosons. To describe spin-2 spinor BEC, we use spin-2 Gross-Pitaevskii Equation (GPE) \cite{pra_ueda_koashi_2002}. Here, we only consider a linear Zeeman term which couples spin and magnetic field. The total energy of spin-2 GPE reads:
\begin{align}
    \nonumber E[\Psi]=&\int \dd^3\mathbf{r}\,\Bigg\{\displaystyle\sum^2_{m=-2}\psi_m^*(\mathbf{r})\Bigg[-\frac{\hbar^2}{2M}\nabla^2+U_{\mathrm{trap}}(\mathbf{r})\\\nonumber&-(-g_n\mu_N)\mathbf{B}(\mathbf{r})\mathbf{\cdot f}+\Omega\hat{L}_z\Bigg]\psi_m(\mathbf{r})\\&+\frac{c_0}2 n^2(\mathbf{r})+\frac{c_1}2|\mathbf{F}(\mathbf{r})|^2+\frac{c_2}2|A_{00}(\mathbf{r})|^2
    \Bigg\}\nonumber\\[2mm]
    &+E_{\mathrm{int}}[\psi_m,\phi_p],\label{All_energy_GPE}
\end{align}
where $\psi_m(\mathbf{r})$ ($m=-2,-1,0,1,2$) denotes a 5-component order parameter, $M_n=2m_n$ is the mass of a neutron Cooper pair, $n(\mathbf{r})=\displaystyle\sum^2_{m=-2}\bigl|\psi_m(\mathbf{r})\bigr|^2$ is the superfluid density, $A_{00}(\mathbf{r})=\frac1{\sqrt{5}}\bigl[2\psi_2(\mathbf{r})\psi_{-2}(\mathbf{r})-2\psi_1(\mathbf{r})\psi_{-1}(\mathbf{r})+\psi^2_0(\mathbf{r})\bigr]$ denotes the amplitude of the spin-singlet pair. The $\mathbf{B\cdot f}$ term describes linear Zeeman effect, where $g_n$ is the $g$-factor for neutrons and $\mu_N$ is the nucler magneton. $\mathbf{f}=(f_x,f_y,f_z)$ where each component of $f_\nu$ $(\nu=x,y,z)$ is $5\times 5$ spin-2 matrix, and $F_\nu=\displaystyle\sum^2_{m,m'=-2}\psi^*_m(\mathbf{r})(f_\nu)_{mm'}\psi_{m'}(\mathbf{r})$. The last term in Eq.~\eqref{All_energy_GPE} represents the energy associated with proton-neutorn interactions, which will be given in Sec.~\ref{Sec:pn-int}, where $\phi_p(\mathbf{r})$ denotes the proton order parameter. $c_0$, $c_1$, and $c_2$ are the coupling constants of the corresponding interactions. $\Omega$ is the rotational velocity of the container.

\subsection{GLE for $^1\text{S}_0$ Proton Cooper Pairs}
\label{sec-3}

In the outer core of neutron stars, protons are considered to form $^1\text{S}_0$ Cooper pairs. Such a proton superconductor may be treated as BEC of spin-0 charged bosons. Here, we use Ginzburg-Landau equation (GLE) to describe the order parameter of proton superconductor, following Refs.~\cite{physik_materie_1966,low_temperature_kopnin_2002,tinkham_superconductivity,mnras_drummond_melatos_2018}. The total energy of spin-0 GLE reads:
\begin{align}
    \nonumber E=&\int \dd^3\mathbf{r}\,\Bigg\{-\frac{\hbar^2}{2M_p}\bigl|(\nabla-i\mathbf{A}(\mathbf{r}))\phi_p(\mathbf{r})\bigr|^2+U_{\mathrm{trap}}(\mathbf{r})\bigl|\phi_p(\mathbf{r})\bigr|^2\\ &+\alpha\bigl|\phi_p(\mathbf{r})\bigr|^2+\frac{\beta}2\bigl|\phi_p(\mathbf{r})\bigr|^4+\phi_p^*(\mathbf{r})(\Omega \hat{L}_z)\phi_p(\mathbf{r})\Bigg\}\nonumber\\[1mm]
    &+E_{\mathrm{int}}[\psi_m,\phi_p],
    \label{GLE}
\end{align}
where $M_p=2m_p$ is the mass of a proton Cooper pair, $\mathbf{A}(\mathbf{r})$ is the vector potential for magnetic field, and $\alpha$ ($<0$) and $\beta$ ($>0$) are the effective coupling constants, respectively.

\subsection{Fluxtube-like External Magnetic Field}

As a first step to investigate the effects of interactions between $^3\text{P}_2$ superfluid vortices and fluxtubes, we introduce a localized magnetic field that mimics a fluxtube. The magnitude of the magnetic flux can be estimated by the London equation \cite{abrikosov}, which leads magnetic flux quanta:
\begin{equation}
    \label{London_eq} \Phi = \frac{h}{2e},
\end{equation}
where $h$ is the Plank's constant and $e$ is the elementally charge. To represent a fluxtube, we consider localized, tube-shaped magnetic field, oriented along the $z$ axis, with a certain radius $r_{\mathrm{FT}}$. The linear Zeeman term in Eq.~\eqref{All_energy_GPE} is given by
\begin{equation}
\label{Zeeman_term_size} g_n\mu_NB(r)=\begin{cases}\displaystyle{\frac{g_n\mu_Nh}{2e(\pi r_{\mathrm{FT}}^2)}}&(r<r_{\mathrm{FT}}),\\[3mm]
0&(r>r_{\mathrm{FT}}).\end{cases}
\end{equation}
% magnetic field

\subsection{Neutron-Proton Interactions}\label{Sec:pn-int}

In GPE and GLE [Eqs.~\eqref{All_energy_GPE} and \eqref{GLE}], to describe nuclear interactions between neutrons and protons, we add an effective interaction term represented as follows:
\begin{align}
\label{effective_interaction} E_{\mathrm{int}}=\int \dd^3\mathbf{r} \left(\displaystyle\sum^2_{m=-2}\xi_m|\psi_m(\mathbf{r})|^2|\phi_p(\mathbf{r})|^2+\displaystyle\sum^2_{m=-2}\eta_m\mathbf{J}_m(\mathbf{r})\cdot\mathbf{J}_p(\mathbf{r})\right),
\end{align}
where the first and the second terms in the parentheses correspond to density-dependent and current-current interactions between neutrons and protons, respectively. This interaction term is almost the same as the one used for the recent analysis of interactions between neutron vortices and proton fluxtubes \cite{mnras_drummond_melatos_2018}. In this work, we disregard $m$-dependence of the coupling constants, setting $\xi_m=\xi$ and $\eta_m=\eta$ for all $m$ values.

\subsection{Computational Setup}

We solve GPE and GLE in real space, discretizing Cartesian coordinates into a uniform grid. {We work with $100\times100$ with a mesh spacing of 0.1\,fm and $50\times50\times50$ grids with a mesh spacing of 0.4\,fm} for two-dimensional (2d) and three-dimensional (3D) simulations, respectively. To confine neutron and proton BECs, we adopt a circular (spherical) shaped external potential for $U_\text{trap}(\mathbf{r})$ in 2D (3D) geometries. To calculate the first- and second-order derivatives, we employ the spectral method, executing the \texttt{FFTW} routines.

To investigate the effects of magnetic field on vortex configurations, in this work, we impose $4\pi$ phase rotation about the center of the trap for the neutron $^3\text{P}_2$ superfluid, while $2\pi$ phase rotation is imposed for the proton $^1\text{S}_0$ superconductor. We introduce a tube-shaped magnetic field that mimics a fluxtube. The magnetic field strength is estimated by Eq.~\eqref{Zeeman_term_size}. To investigate the shape of quantum vortices in the ground state, we solve the GPE and GLE using the imaginary-time evolution method.

\section{Computational Results}
\label{results}

\subsection{Magnetic-field effects on Vortex Shape in the 2D Case}
% 2D-> m=0

First, we report an illustrative example of a 2D simulation, showing possible effects of magnetic field on the vortex structure of $^3\text{P}_2$ superfluid.

\begin{figure}[t]
    \centering
    Without magnetic field\\[-0.2mm]
    \begin{subfigure}[b]{0.47\linewidth}
    \includegraphics[width=\linewidth]{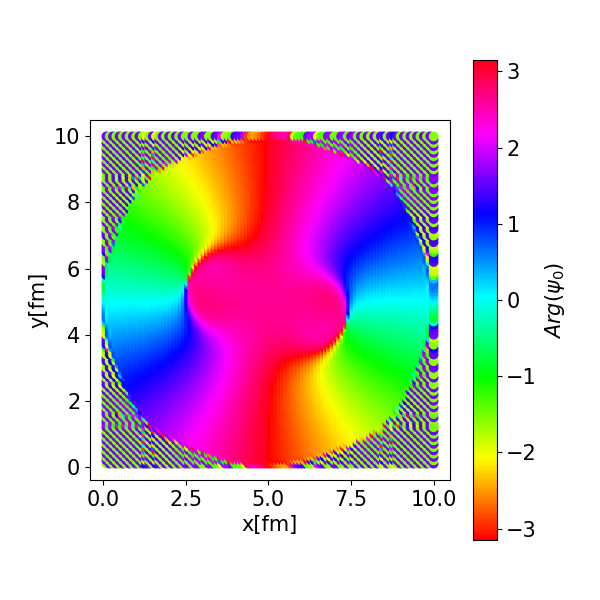}\vspace{-6mm}
    \caption{Arg$\psi_0$}
    \label{2d_noint_arg0}
    \end{subfigure}
    \begin{subfigure}[b]{0.47\linewidth}
    \includegraphics[width=\linewidth]{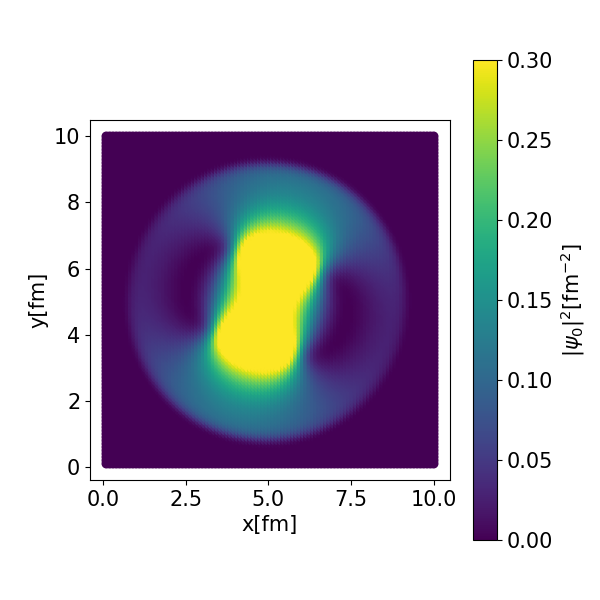}\vspace{-6mm}
    \caption{$|\psi_0|^2$}
    \label{2d_noint_rho0}
    \end{subfigure}\vspace{-3mm}
    With magnetic field\\[-0.3mm]
    \begin{subfigure}[b]{0.47\linewidth}
    \includegraphics[width=\linewidth]{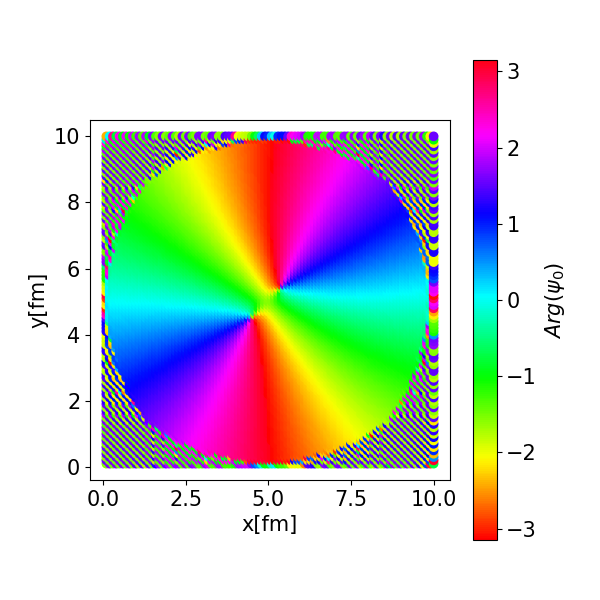}\vspace{-6mm}
    \caption{Arg$\psi_0$}
    \label{2d_mag_arg0}
    \end{subfigure}
    \begin{subfigure}[b]{0.47\linewidth}
    \includegraphics[width=\linewidth]{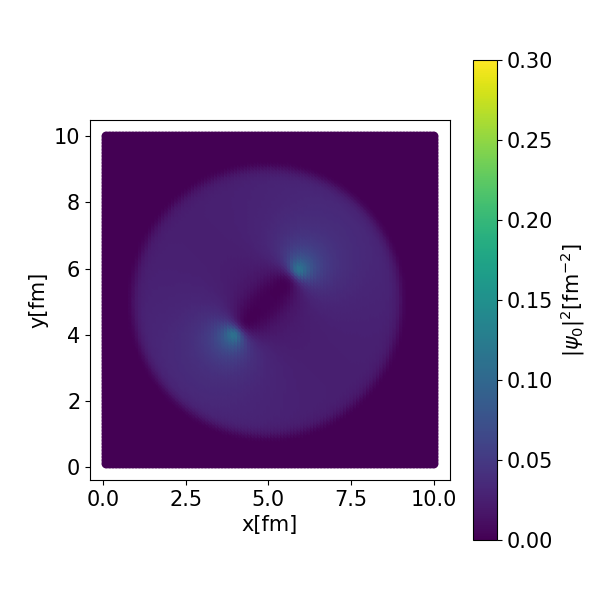}\vspace{-6mm}
    \caption{$|\psi_0|^2$}
    \label{2d_mag_rho0}
    \end{subfigure}\vspace{-4mm}
    \caption{
    Results of 2D calculations for the case with two or four $^3\text{P}_2$ vortices (see texts for details) and a single fluxtube-like magnetic field. Figures \ref{fig:2d_mag_0_vortex_change}(a) and \ref{fig:2d_mag_0_vortex_change}(b) show phase and density of $\psi_0$ components of $^3P_2$ superfluid without any interaction with the fluxtube, while Figs.~\ref{fig:2d_mag_0_vortex_change}(c) and \ref{fig:2d_mag_0_vortex_change}(d) are those with an interaction with the fluxtube. In this case, we neglect the proton-neutron interaction: \textit{i.e.}, $\xi=0$ and $\eta=0$ in Eq.~\eqref{effective_interaction}.
    }\vspace{-8mm}
    \label{fig:2d_mag_0_vortex_change}
\end{figure}

In Fig.~\ref{fig:2d_mag_0_vortex_change}, we show the phases [(a) and (c)] and the density [(b) and (d)] of the $m=0$ component of neutron superfluid. Panels (a) and (b) correspond to the results obtained without magnetic field, while panels (c) and (d) correspond to those with a tube-like external magnetic field at the center. As can be inferred from Fig.~\ref{fig:2d_mag_0_vortex_change}(a), there are four half-integer quantum vortices (HQVs) of the $m=0$ component. On the contrary, when we apply the magnetic field, the HQVs are not formed, but two integer quantum vortices (IQVs) are observed.

\begin{figure}[t]
    \centering
    \includegraphics[width=0.9\linewidth]{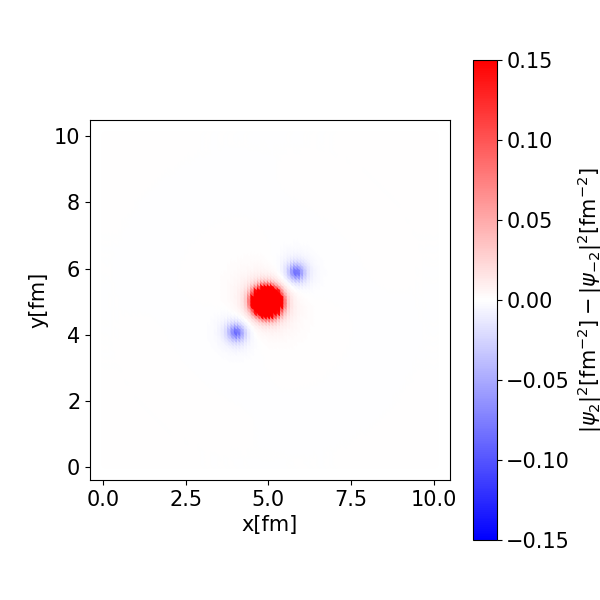}\vspace{-8mm}
    \caption{2D calculation of $^3P_2$ vortices and a tube-like magnetic field. Spin polarization of neutron makes $|\psi_2|$ and $|\psi_{-2}|$ density differences.
    Here we neglect proton-neutron interaction term:$\xi=0,\eta=0$.}\vspace{-8mm}
    \label{fig:2d_mag_dif2m2}
\end{figure}

Actually, the observed change of $m=0$ vortex configurations is caused by the polarization of the other components ($m=\pm1,\pm2$). To demonstrate this point, in Fig.~\ref{fig:2d_mag_dif2m2}, we show spin polarization, defined as the difference $|\psi_2|^2-|\psi_{-2}|^2$ in the $x$-$y$ plane. From the figure, it is clearly visible that the $m=2$ component is accumulated inside the magnetic-field region, exhibiting substantial polarization at the center. Intriguingly, as a side effect of this polarization, the $m=-2$ component is accumulated inside the two neutron vortices. In this way, the presence of magnetic field breaks the symmetry between the $m=\pm2$ components, allowing for the $m=\pm2$ components to form completely different distributions as shown Fig.~\ref{fig:2d_mag_dif2m2}. Due to these distributions of $\psi_2$ and $\psi_{-2}$, the density distribution of $\psi_0$ is affected by the spin mixing term, $|\mathbf{F}|^2$ and $|A_{00}|^2$, resulting in the transition from 4 HQVs to 2 IQVs.

\subsection{Magnetic-field effects on Vortex Shape in the 3D Case}

\begin{figure}
    \centering\vspace{-10mm}
    \includegraphics[width=\linewidth]{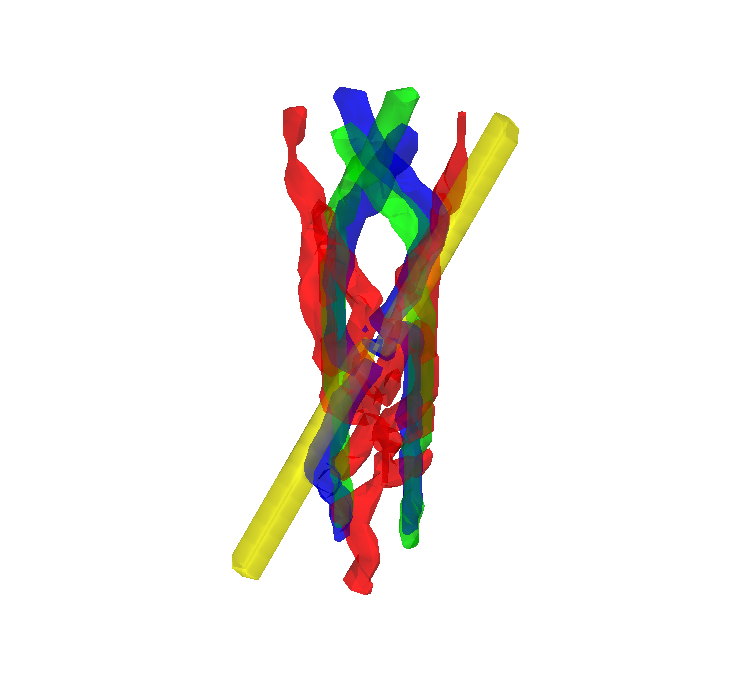}\vspace{-11mm}
    \caption{A result of 3D calculation for $^3\text{P}_2$ vortices and a fluxtube-like magnetic field (indicated by a yellow tube). Red, blue, green isosurfaces indicate vortex core positions of $\psi_0$, $\psi_2$, and $\psi_{-2}$ components, respectively. In this case, we neglect the proton-neutron interaction: \textit{i.e.}, $\xi=0$ and $\eta=0$.}\vspace{-8mm}
    \label{fig:3d_mag_1}
\end{figure}
% 3D->

Next, we report a tentative result of 3D simulations. A remarkable difference between 2D and 3D geometries is that, in the latter case, vortices can bend and entangled each other, and they are not necessarily aligned along the fluxtube direction.

To discuss shape changes of quantum vortices due to the presence of a magnetic fluxtube, we need to define vortex-core positions in 3D space. To this end, we analyze the change in the phase by performing the line integral on the eight grid points around each point. If the phase rotates by $\pm2\pi$, we regard it as a single IQV at that point. If the phase rotate by $\pm\pi$, we regard it as a single HQV. 

To show how a fluxtube affects the vortices configuration in 3D, we show in Fig.~\ref{fig:3d_mag_1} vortices of $m=0$ (red), $m=2$ (blue), and $m=-2$ (green) components, in the presence of a fluxtube (yellow). The vortices of $m=2$ (blue) and $m=-2$ (green) are in different positions near the fluxtube. Because of the polarization effect discussed in the 2D case as well, $m=2$ vortices (blue) are attracted by the fluxtube (yellow). This polarization affects the vortex position of $m=0$ component (red).

As demonstrated by these calculations, there is a non-trivial complex interplay between neutron-proton interactions and spin magnetic-field interactions and associated topological structure of $^3\text{P}_2$ and $^1\text{S}_0$ order parameters. We thus consider that it is important to carefully examine vortex-fluxtube configurations in various situations to explore possible impacts on physics of neutron stars.

\vspace{-2mm}
\section{Summary and Perspectives}

Pulsar glitches are considered as a macroscopic astrophysical phenomenon caused by dynamics of microscopic quantum vortices that offers a unique opportunity to explore superfluid and superconducting properties in dense nuclear matter. In this work, we focus on the outer core region where neutron $^3\text{P}_2$ superfluid vortices and proton $^1\text{S}_0$ fluxtubes coexist, which have not been well studied to date. In this contribution, we have reported our ongoing attempt to microscopically describe vortex-fluxtube interactions and dynamics, showing that $^3\text{P}_2$ vortex configurations could be affected substantially by fluxtubes.

As a next step, we plan to determine the magnetic field inside the fluxtube in a self-consistent manner and analyze the vortex-fluxtube dynamics through time-dependent simulations. By solving the magnetic field configuration self-consistently, we can, not only analyze the effects on vortices in situations where multiple fluxtubes exist, but also investigate how vortices affect on the arrangement of fluxtubes. Furthermore, by tracking time evolution in rotating systems, we will analyze how the magnetic field structure associated with fluxtubes is affected by the $^3\text{P}_2$ vortices in the outer core region.

\vspace{-2mm}
\section*{Acknowledgments}

This work is supported by JSPS Grant-in-Aid for Scientific Research, Grants No.~23K03410, No.~23K25864, and No.~JP25H01269, and by JST SPRING, Japan, Grant Number JPMJSP2180. This work used computational resources of the  Yukawa-21 supercomputer provided by YITP, Kyoto University.

%
% BibTeX or Biber users please use (the style is already called in the class, ensure that the "woc.bst" style is in your local directory)
% \bibliography{your_bib_file} % Replace "your_bib_file" with the actual name of your .bib file

\end{document}